\documentclass[superscriptaddress,prx,twocolumn]{revtex4}
\usepackage{graphicx}
\usepackage{amsmath,physics}
\usepackage{color}
\usepackage{hyperref}
\graphicspath{ {final_images/} }
\usepackage{ulem}

\begin{document}
\title{An adaptive variational algorithm for exact molecular simulations on a quantum computer}
\author{Harper R. Grimsley}
\affiliation{Department of Chemistry, Virginia Tech, Blacksburg, VA 24061, USA}
\author{Sophia E. Economou}
\affiliation{Department of Physics, Virginia Tech, Blacksburg, VA 24061, USA}
\author{Edwin Barnes}
\affiliation{Department of Physics, Virginia Tech, Blacksburg, VA 24061, USA}
\author{Nicholas J. Mayhall}
\email{nmayhall@vt.edu}
\affiliation{Department of Chemistry, Virginia Tech, Blacksburg, VA 24061, USA}

\begin{abstract}
\section*{Abstract}
	\textbf{Quantum simulation of chemical systems is one of the most promising near-term applications of quantum computers. The variational quantum eigensolver, a leading algorithm for molecular simulations on quantum hardware, has a serious limitation in that it typically relies on a pre-selected wavefunction ansatz that results in approximate wavefunctions and energies. Here we present an arbitrarily accurate variational algorithm that instead of fixing an ansatz upfront, this algorithm grows it systematically one operator at a time in a way dictated by the molecule being simulated. This generates an ansatz with a small number of parameters, leading to shallow-depth circuits. We present numerical simulations, including for a prototypical strongly correlated molecule, which show that our algorithm performs much better than a unitary coupled cluster approach, in terms of both circuit depth and chemical accuracy. Our results highlight the potential of our adaptive algorithm for exact simulations with present-day and near-term quantum hardware.}
\end{abstract}
\maketitle

\section{Introduction}
Anticipation that a useful quantum computer will be realized in the near future
	has motivated intense research into developing quantum algorithms which can potentially make progress on classically intractable computational problems. 
	While many research areas expect to see transformative change with the development of such quantum devices, 
		computational chemistry is poised to be among the first domains to significantly benefit from such new technologies.
	Due to the exponential growth in the size of the Hilbert space with increasing orbitals, 
		a quantum computer with tens of qubits could potentially surpass classical algorithms \cite{Aspuru-Guzik2005,McArdle2018a,Cao2018}. Achieving such a capability depends not only on the quality of the qubits, but also critically on the efficiency of the algorithms.
	
	The phase estimation algorithm (PEA) \cite{Kitaev1995} was the first algorithm proposed for simulating electronic structure problems on a quantum computer \cite{Lloyd1996,Aspuru-Guzik2005}.
	PEA provides a path for obtaining the exact ground state electronic energy for a molecule by evolving in time a quantum state with significant overlap with the ground state using the molecular Hamiltonian of interest.
	Due to the very long circuit depths and complex quantum gates required by PEA, 
		the coherence times needed to simulate interesting electronic states would exceed the coherence times available on any existing or near-term quantum device. Improvements to PEA still require significant resources and experimental demonstrations to date only involve a few qubits \cite{Lanyon2010,OMalley2016,Paesani2017}. 

	In order to reduce the significant hardware demands required by PEA and exploit the capabilities of noisy intermediate-scale quantum (NISQ) devices \cite{Preskill2018}, 
		the variational quantum eigensolver (VQE) algorithm was proposed and demonstrated using photonic qubits by Peruzzo et al. \cite{Peruzzo2014}. 
		This was followed by several theoretical studies on VQE \cite{Whitfield2011,McClean2016,OMalley2016,McClean2017,Barkoutsos2018,Romero2018,Colless2018,Lee2018} and demonstrations 
		on other hardware such as superconducting qubits \cite{OMalley2016,Kandala2017,Colless2018} and trapped ions \cite{Shen2017,Hempel2018}. 
		 Other approaches have been pursued as well,  including methods for adiabatic quantum computation \cite{Xia2017} and quantum machine learning \cite{Xia2018}.

	VQE is a hybrid quantum-classical algorithm, because the computational work is shared between classical and quantum hardware. VQE starts with an assumption about the form of the target wavefunction. Based on this form, an ansatz with several tunable parameters is constructed, and a quantum circuit capable of producing this ansatz is designed. The ansatz parameters are variationally adjusted until they minimize the expectation value of the molecular Hamiltonian. Classical hardware is used to precompute all the Hamiltonian terms and to update the parameters during the circuit optimization. The quantum hardware is only used to prepare a state (defined by its current set of ansatz parameter values) and to perform measurements of the various interaction terms in the molecular Hamiltonian, $\hat{H} = \sum_i g_i \hat{o}_i$. Because the individual operator terms, $\hat{o}_i$, generally do not commute, the state preparation has to be repeated multiple times, until all the individual operators have been measured enough times to get sufficient statistics on their mean value. Details on all these steps can be found in Ref.~\cite{McClean2016}.
	
	Compared to PEA, VQE is much more suitable for NISQ devices, trading in the long circuit depths for shorter state preparation circuits, at the expense of a much higher number of measurements. Although VQE has been demonstrated to be more efficient and error-tolerant \cite{OMalley2016,Colless2018,McClean2016}, this comes with the compromise that the ansatz generally only allows one to obtain approximations to the ground state. Because the choice of ansatz determines the variational flexibility of the trial state, the quality of a VQE simulation is only as good as the ansatz. 
	
	Several approaches have been explored with the goal of creating a compact ansatz which provides high accuracy with few parameters and shallow circuits.
	The first ansatz explored \cite{Peruzzo2014} was based on the unitary variant of coupled cluster theory truncated at single and double excitations (UCCSD), inspired by early efforts in computational chemistry to improve coupled cluster theory \cite{Bartlett1989,Kutzelnigg1991a,Taube2006,Harsha2018a}. In UCCSD, trial states are generated by applying to a reference state a unitary operator in the form of an exponential of a sum of single and double fermion operators with their coefficients taken as free parameters.
	More recent proposals based on UCCSD include the unitary Bogoliubov coupled cluster theory which takes a generalized Hartree-Fock (HF) state as the reference \cite{Dallaire-Demers2018a}
		and the $k$-UpCCGSD approach of Lee et al. \cite{Lee2018} which uses $k$ products of unitary paired generalized doubles excitations, along with the full set of generalized single excitations.
	The $k$-UpCCGSD approach builds on early work by Nakatsuji \cite{Nakatsuji2005,Nakatsuji2000,Nakatsuji2002,Nakatsuji2001} and Nooijen \cite{Nooijen2000a} studying the use of generalized excitation terms 
		in classical quantum chemistry algorithms, but prunes the expansive operator list by restricting the two-particle terms to only paired interactions,  which provides a systematic way to converge to FCI without introducing higher excitation rank operators. 
	Ryabinkin et al.  \cite{Ryabinkin2018} recently proposed a coupled cluster-like ansatz which is constructed directly in the qubit representation with the goal of achieving shallower circuits.
	While not directly a variation of the UCCSD ansatz itself, Ref. \cite{Colless2018} developed an approach  (termed the quantum subspace expansion) to extract not just the expectation value of $\hat{H}$, 
		but all the matrix elements $\bra{I}\hat{H}\ket{J}$ in a small subspace consisting of single excitations from the trial state.
	This Hamiltonian matrix is then diagonalized on a classical computer, which reduces the impact of decoherence and gives access to excited states.
	Even further from the original UCC ansatz, Kandala et. al. \cite{Kandala2017} have used an alternative ansatz for their VQE experiments based on the native entangling gate in their superconducting qubit device, referred to as a ``hardware-efficient ansatz''. This allows entanglement to be created directly from a device-wide unitary instead of through a more traditional gate decomposition of a fermionic operator.

Despite these considerable improvements to the UCCSD ansatz for VQE, this remains an approximate approach that works best for systems that are not strongly correlated. However, strongly correlated systems are the hardest to simulate classically, and this is precisely the motivation for performing simulations using quantum computers. While an exact VQE simulation could in principle be performed by adding higher rank excitations to the ansatz, this would be prohibitively expensive for both the classical subroutines and NISQ devices. To overcome these challenges, we need to avoid imposing an ad hoc ansatz and instead allow the system to determine its own compact, quasi-optimal ansatz.

	In this paper, we achieve this by introducing a simple algorithm termed Adaptive Derivative-Assembled Pseudo-Trotter ansatz Variational Quantum Eigensolver (ADAPT-VQE). ADAPT-VQE
	determines a quasi-optimal ansatz with the minimal number of operators for a desired level of accuracy. 
	The key idea is to systematically grow the ansatz by adding fermionic operators one-at-a-time, such that the maximal amount of correlation energy is recovered at each step.
	This results in a wavefunction ansatz that is discovered by the algorithm, and which cannot be predicted a priori from a traditional excitation-based scheme like UCCSD.
	While intuitive, this approach can also be derived more rigorously as a particular optimization procedure for Full Configuration Interaction (FCI) VQE and is more thoroughly discussed in the Section 1 of the Supplement. We demonstrate the power of ADAPT-VQE through numerical simulations of three molecules of increasing complexity: LiH, BeH$_2$, and H$_6$. In each case, we find vastly improved performance compared to UCCSD, both in terms of the number of operators needed to form the trial states and in terms of chemical accuracy. Therefore, we believe that ADAPT-VQE is an ideal hybrid algorithm for NISQ devices.

\section{Results}
\subsection{Specification of the adopted notation}
In order to define the approach, several definitions and notations need to be established. 
First, molecular orbital indices $i$ and $j$ denote occupied orbitals, $a$ and $b$ denote virtual orbitals, and $p$, $q$, $r$ and $s$ denote arbitrary molecular orbitals. 
In coupled cluster theory, in particular CCSD, an expansion based on the HF state $\ket{\psi^\text{HF}}$ is created by using an exponential ansatz involving single and double excitation operators:
\begin{align}
	\ket{\psi^\text{CCSD}} =&  e^{\hat{T}_1+\hat{T}_2}\ket{\psi^\text{HF}},
\end{align}
where the excitation operators are defined as:
\begin{align}
	\hat{T}_1 &= \sum_{ia} \hat{t}_i^a  = \sum_{ia} t_i^a \hat{a}^\dagger_a\hat{a}_i, \\
	\hat{T}_2 &= \sum_{i\textless j, a\textless b} \hat{t}_{ij}^{ab} = \sum_{i\textless j, a\textless b} t_{ij}^{ab}\hat{a}^\dagger_a\hat{a}^\dagger_b\hat{a}_i\hat{a}_j .
\end{align}
For closed shell molecules near equilibrium, CCSD provides a robust ansatz for molecular simulations. 
Early efforts to combine size extensivity and variationality were pioneered by Bartlett, Kutzelnigg, and coworkers \cite{Bartlett1989,Kutzelnigg1991a,Taube2006}.
In this context, a unitary variant of coupled cluster theory (UCCSD) was defined by replacing the excitation operators with an anti-Hermitian sum of excitation and de-excitation operators:
\begin{align}
	\hat{t}_{ij}^{ab} \rightarrow  \hat{t}_{ij}^{ab}-\hat{t}_{ab}^{ij} = \hat{\tau}_{ij}^{ab}.
\end{align}
Because UCCSD is based on a unitary operator, the adjoint is the inverse, and the expectation value of the UCCSD wavefunction can be expanded using the Baker-Campbell-Hausdorff (BCH) formula to obtain a normalized Hamiltonian expectation value (Rayleigh quotient) for variational optimization. 
Unfortunately, the BCH expansion does not truncate at finite order, making UCCSD computationally intractable on classical hardware. 
However, the unitary nature of UCCSD is actually a benefit for quantum algorithms as it corresponds to a coherent time evolution, and this was the original motivation for using UCCSD in VQE \cite{Peruzzo2014}.

In addition to a unitary form, CCSD can also be generalized by including excitation operators which immediately annihilate the HF state. 
These would include excitations from occupied to occupied, virtual to virtual, etc. 
Generalized excitations or interactions of this form have been considered prevously, and have been used in the context of VQE recently by Lee and coworkers \cite{Lee2018}.
In this case the cluster operators are further generalized to remove the HF-based subspace restriction: $\hat{\tau}_{ij}^{ab} \rightarrow  \hat{\tau}_{pq}^{rs}$,
where $p$, $q$, $r$, and $s$ refer to any arbitrary orbital. 

Although UCCSD is perhaps a natural ansatz for VQE, it cannot be implemented directly as written or as explored previously in the quantum chemistry context. 
Because the gate model of quantum computation is realistically bound to using gates acting on only a  few qubits at a time, 
	the UCCSD operator must be broken up into a time-ordered sequence of few (one or two)  particle operators. 
This is achieved by using a Trotter expansion of a matrix exponential \cite{Hatano2005},
\begin{align}
	e^{A+B} = \lim_{n\rightarrow \infty} \left(e^{(A/n)}e^{(B/n)}\right)^n.
\end{align}
 Because the generalized single and double excitation operators do not commute, the use of a truncated Trotter expansion represents an approximation to the underlying UCCSD ansatz, 
	and recent work has shown clearly that this does not strongly affect the results because the variational flexibility is sufficient to absorb this error \cite{OMalley2016},
	and that even a single Trotter number ($n=1$) is sufficient to reproduce the results of UCCSD.
As a result, a unitary, generalized, Trotterized ansatz becomes:
\begin{align}
	\ket{\psi^\text{tUCC}} =& \prod_{s\in\{pq\}} e^{\hat{t}_s} \prod_{d\in\{pqrs\}} e^{\hat{t}_d} \ket{\psi^\text{HF}},
\end{align}
where notation is introduced such that the generalized singles index, $s$, runs over all unique pairs of $p,q$ and the doubles index, $d$,  over unique combinations of $p,q,r,s$. 

\begin{figure*}
	\includegraphics[width=\linewidth]{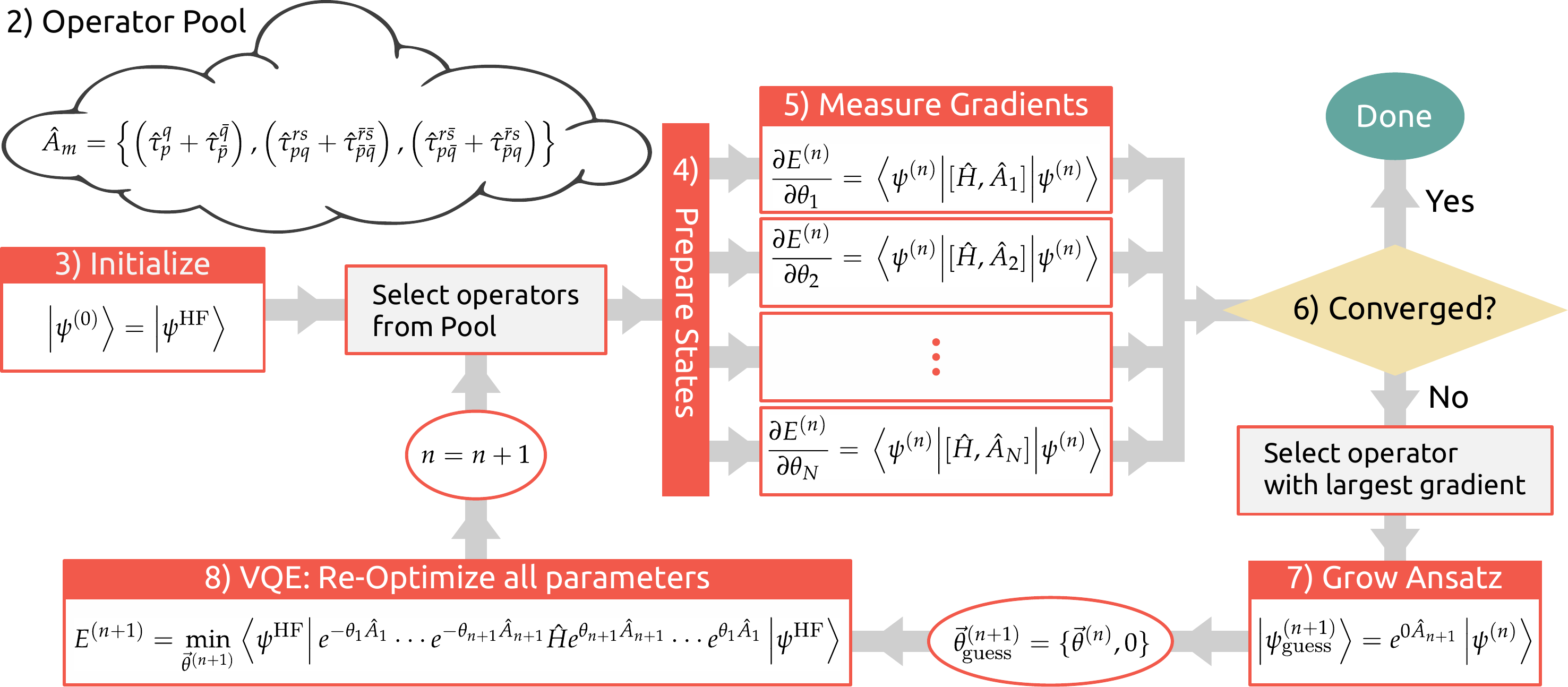}
	\caption{Schematic depiction of the ADAPT-VQE algorithm described presented.
	 Since step 1 occurs on classical hardware, it is not included in the illustration.
	$\vec{\theta}^{(n)}$ is the list of ansatz parameters at the $n$th iteration.
	The number of parameters, $len(\vec{\theta}^{(n)})$, is equal to the number of operators in the ansatz. 
	``Operator Pool'' refers to the collection of operators which are used to grow the ansatz one-at-a-time. 
	Each $\hat\tau_p^q$ represents a generalized single or double excitation, 
	and these operators are then spin-complemented.
	The orbital indices refer to spatial orbitals, and the overbar indicates $\beta$ spin.
	Orbital indices without overbars have $\alpha$ spin. 
	Note that growing the ansatz does not drain the pool, and so operators can show up multiple times if selected by the algorithm.
	} \label{fig:algorithm}
\end{figure*}

\subsection{ADAPT-VQE algorithm}\label{sec:adaptvqe}
The above discussion described the Trotter expansion as an approximation to UCCGSD.
However, as recognized previously \cite{OMalley2016,Barkoutsos2018a},
	if the parameters are optimized after the Trotterization, this is not so much an approximation to UCC as it is a wholly unique ansatz.
In fact, the exact FCI solution could be obtained by simply going to an $n$th order Trotterized form of UCCSD
	and allowing the different parameter replicas to vary independently. 
This is due to the fact that $n$-body interactions can be described as products of one- and two-body interactions.  
The exact (FCI) quantum state can thus be represented as an arbitrarily long product of one- and two-body operators, 
\begin{align}\label{eq:fcivqe}
	\ket{\psi^\text{FCI}} = \prod_k^\infty\prod_{pq}e^{\hat{\tau}_p^q(k)}\prod_{pqrs}e^{\hat{\tau}_{pq}^{rs}(k)}\ket{\psi^\text{HF}},
\end{align}
where $\hat{\tau}_{pq}^{rs}(k)$ is the $k$th instance, or ``replica", of the operators in $\hat{t}_{pq}^{rs} - \hat{t}_{rs}^{pq}$. It is important to note that this is not a Trotter approximation to any simple two-body ansatz, as each replica can assume different parameter values, 
	e.g., $\tau_{pq}^{rs}(k) \neq \tau^{rs}_{pq}(j)$.

The main goal in this paper is to approximate FCI with arbitrary accuracy using a maximally compact sequence of unitary operators. 
The basic outline of the algorithm is drawn schematically in Fig. \ref{fig:algorithm} and is as follows:
\begin{enumerate}
	\item On classical hardware, compute one- and two-electron integrals, and transform the fermionic Hamiltonian into a qubit representation using an appropriate 
		transformation: Jordan-Wigner, Bravyi-Kitaev, etc. This is a standard step in regular VQE.
	\item Define an ``Operator Pool''. This is simply a collection of operator definitions which will be used to construct the ansatz.  
		For the examples presented in the next section, we consider the set of all unique spin-complemented one- and two-body operators,	 
		but one might imagine adding a few three-body or four-body terms as well. 
	\item Initialize qubits to an appropriate reference state, ideally one with the correct number of electrons.
		The HF state would be a sensible choice here. Initialize the ansatz to the identity operator. 
	\item On a quantum computer, prepare a trial state with the current ansatz.
		If multiple quantum computers are available, perform this step on all devices simultaneously.
	\item Measure the commutator of the Hamiltonian with each operator in the pool to get the gradient.
		Repeating this multiple times and averaging gives the gradient of the  expectation value of the Hamiltonian with respect to the coefficient of each operator.
		This can be done in parallel.
	\item If the norm of the gradient vector is smaller than some threshold, $\epsilon$, exit. 
	\item Identify the operator with the largest gradient and add this single operator to the left end of the ansatz, with a new variational parameter. 
		Note that this does not ``drain'' the pool in the sense that choosing an operator does not remove it from the pool so it can be used again later.
	\item Perform a VQE experiment to re-optimize all parameters in the ansatz. 
	\item Go to step 4.
%
\end{enumerate}

As described above and illustrated in Fig. \ref{fig:algorithm}, each iteration starts as a series of uncoupled experiments to obtain the parameter gradients via measurements of operator commutators (the gradient expression in step 5 is derived in section IB of the Supplemental Information).
The purpose of these gradient measurements is to determine the best operator with which to grow the ansatz, 
	as the operator with the largest gradient is likely to recover the most correlation energy in the subsequent VQE minimization. 
This process is continued iteratively, until a convergence threshold is met.
In the classical numerical examples presented below, we chose to consider the $L^2$ norm of the gradient vector to determine convergence. This is just one possibility, and alternative convergence indicators could be used instead in step 6. At convergence, the ADAPT-VQE algorithm obtains the following ansatz:
\begin{align}
	\ket{\psi^{\text{ADAPT}(\epsilon)}} = \left(e^{\hat{\tau}_N}\right)\left(e^{\hat{\tau}_{N-1}}\right)\cdots\left(e^{\hat{\tau}_{2}}\right)\left(e^{\hat{\tau}_{1}}\right)\ket{\psi^\text{HF}}
\end{align}
where the identity of each $\hat{\tau}_i$ is determined by the algorithm.

The re-optimization subroutine in step 8 can be implemented on either a classical or quantum processor using any of the gradient- or non-gradient-based optimization routines that have been proposed or demonstrated for VQE \cite{McClean2016,Kandala2017,Colless2018,Romero2018}. Note that this subroutine is distinct from the gradient computed in step 5 of the algorithm. Additional possible modifications to the algorithm are mentioned in the Discussion.

The evaluation of all the gradient terms could in principle be achieved in a NISQ-friendly, highly parallel 
	manner with a large number of uncoupled quantum computers all tasked with preparing the same state and measuring a different operator.
This is the same potential for parallelization that the underlying VQE subroutine has.
Just as with the original motivation for VQE, ADAPT-VQE decreases the circuit depth at the expense of a larger number of measurements. 
In our case a sequence of VQE experiments is performed, with the most resource-demanding experimental steps happening at the end. 
This constitutes a rather large prefactor which would scale with the size of the system, 
	but the crucial advantage is controllability over the ansatz accuracy (in principle approaching FCI).
Because the number of non-zero parameters equals the number of iterations, in order to discover an ansatz for a large system,
an equal number of VQE re-optimizations will need to be performed. 
One strategy to minimize this prefactor could simply be to add a few operators at a time. 

Determining resource requirements for adaptive procedures is rather difficult. 
The classical resources are not expected to be significant in the foreseeable future. 
However, as quantum technology progresses toward deeper circuits, the parameter manipulation and updating on a classical computer could become costly. However, we expect the dependence between parameters at the beginning and end of the ADAPT-VQE circuit to decay with circuit depth, 
	such that one could imagine freezing the early parameters after a certain number of iterations. 
This would possibly establish an approach for FCI with only a polynomial number of variables,
	completely avoiding any exponential cost for the classical hardware. 
\begin{figure*}
	\includegraphics[width=\linewidth]{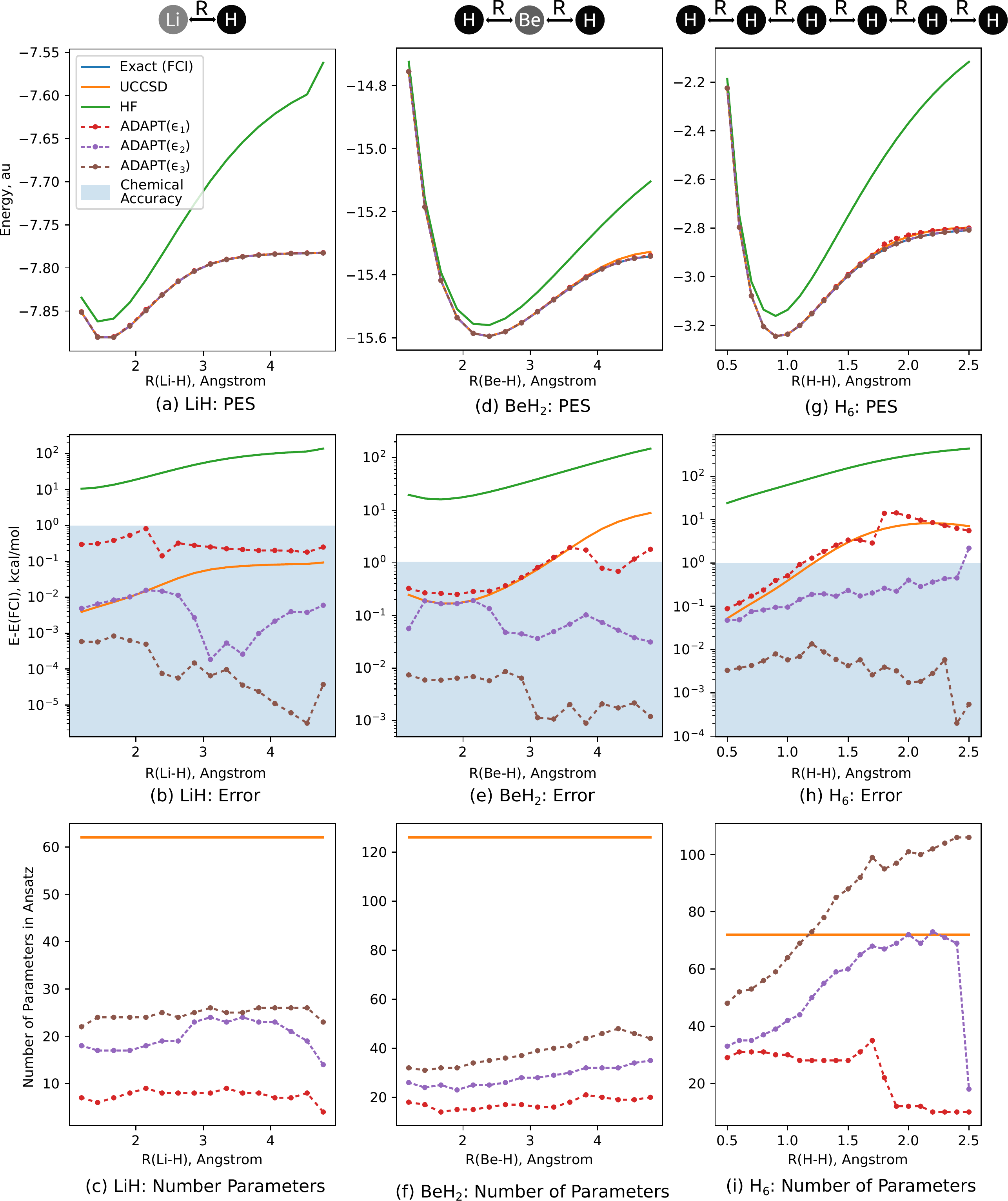}
	\caption{Dissociation curves for LiH, BeH$_2$, and H$_6$. Potential energy as a function of nuclear coordinate, Hartree units (a, d, g).  
	Absolute energy differences from FCI, kcal/mol units (b, e, h). Shaded blue region represents area within ``chemical accuracy'' as 1 kcal/mol.
	Number of variational operators in associated ansatz (c, f, i). 
	Notation: $\epsilon$ indicates gradient norm threshhold used such that $\epsilon_m = 10^{-m}$.
	In all curves, the FCI curve lies directly underneath the ADAPT($\epsilon_3$) curve, and so is not visible.
	} \label{fig:pes}
\end{figure*}

\subsection{Molecular dissociation simulation results}\label{sec:numex}
In this section, we explore the convergence properties of the ADAPT-VQE algorithm with a few small molecular systems,
LiH, BeH$_2$, and linear H$_6$. The former two molecules have been simulated using quantum hardware \cite{Kandala2017,Hempel2018}.
 H$_6$ is included as a prototypical strongly correlated molecule,
which allows us to test the ADAPT-VQE approach for systems which are not well described with unitary coupled cluster. 

In order to perform the simulations, an in-house code was written, using Psi4 \cite{Turney2012a,Smith2018} for the integral calculation (via the OpenFermion-Psi4 \cite{McClean2017a} interface)
and OpenFermion was used for the Jordan-Wigner operator transformation. 
All calculations used the Broyden-Fletcher-Goldfarb-Shannon (BFGS) minimization implemented within Scipy \cite{scipy}.
To classically simulate the re-optimization subroutine in step 8 of ADAPT-VQE, we could use a standard numerical gradient method. 
However, in order to improve the efficiency and allow precise gradients for tight convergence, we derived and implemented an efficient analytic gradient function, which is detailed in Section D of the Supplement. By reusing intermediates between individual parameter gradients, 
	this algorithm obtains the full gradient vector for all parameters at a cost which is only roughly 2x that of the base energy evaluation. 
A table with timing data is included in Table 1 of the Supplement.

As discussed in the previous section, the ADAPT ansatz uses a convergence threshold to determine when the calculation should terminate (step 6). Here we use the norm of the gradient vector and compare it to threshold $\epsilon_m$, which we define as
\begin{align}
\epsilon_m = 10^{-m}.
\end{align}
For example, an ADAPT-VQE calculation where the norm of the operator pool gradient is converged to less than 0.001 would be denoted as ADAPT($\epsilon_3$). In what follows we present numerical results for bond-dissociation curves for LiH, BeH$_2$ and H$_6$ for three different choices of the threshold ($m=1,2,3$). We also investigate alternate protocols for the ansatz growth and demonstrate the superiority of the ADAPT ansatz.

Here, 
we study the LiH bond dissociation computed using several methods, including FCI, UCCSD (un-Trotterized), HF, ADAPT($\epsilon_1$), ADAPT($\epsilon_2$), and ADAPT($\epsilon_3$), all with the STO-3G basis set. In this basis set, LiH has 6 spatial orbitals and a Hilbert space of dimension 4096. By starting with the HF state with two $\alpha$ (spin-up) and two $\beta$ (spin-down) electrons and using only number conserving operators,
the relevant subspace to explore has a dimension of ${6 \choose 2}\cdot{6 \choose 2} = 225$. In this basis, the occupied orbitals are \{1,2\}, and the virtual orbitals are \{3,4,5,6\}.

The bond dissociation curves are shown in Fig.~\ref{fig:pes}(a), where all the curves, with the exception of HF, cannot really be distinguished on this scale. However, as shown in Fig.~\ref{fig:pes}(b), when the FCI energy is subtracted and the scale is adjusted, significant differences become evident. Shading is used to indicate chemical accuracy, which is achieved in all cases other than HF. LiH has only a single pair of electrons (a $\sigma$ bond) breaking along the dissociation coordinate, and UCCSD exhibits chemical accuracy throughout the curve. While ADAPT($\epsilon_1$) is not as accurate as UCCSD, ADAPT($\epsilon_2$) is comparable to UCCSD at short bond distances and comfortably outperforms it at longer distances. This is also evident in Table~\ref{tbl:pes}, where the average error across the potential energy surface (PES) is shown. Remarkably, ADAPT($\epsilon_3$) outperforms UCCSD  throughout the whole curve by at least an order of magnitude and in some cases up to four orders of magnitude. 

Even more impressive is how few parameters are needed to achieve this level of accuracy. As shown in Fig.~\ref{fig:pes}(c), in all three cases and for all bond distances, ADAPT is much more compact than UCCSD. UCCSD has 92 parameters, which can be reduced to 64 by combining spin-complements. In all three ADAPT calculations, fewer than half of the parameters are needed compared to UCCSD. Although UCCSD is noticeably more accurate than the simplest ADAPT calculation with a gradient norm threshold of 0.1, the ADAPT($\epsilon_1$) ansatz is incredibly compact, consisting of fewer than 10 parameters across the curve.
For example, the ADAPT($\epsilon_1$) ansatz for LiH at bond distance 2.39 \AA{} is
\begin{align}
	\big|\psi^{\text{ADAPT}(\epsilon_1)}\big\rangle =   e^{\hat{\tau}_{12}^{16}}   e^{\hat{\tau}_{2\bar{2}}^{5\bar{5}}} 
		e^{\hat{\tau}_{2\bar{2}}^{4\bar{4}}} e^{\hat{\tau}_{12}^{13}} e^{\hat{\tau}_{1\bar{1}}^{3\bar{3}}}  
		e^{\hat{\tau}_{2\bar{2}}^{3\bar{3}}} e^{\hat{\tau}_{2\bar{2}}^{3\bar{6}}} e^{\hat{\tau}_{2\bar{2}}^{6\bar{6}}}\ket{\psi^\text{HF}},\label{LiHansatz}
\end{align}
which includes a mixture of both double excitations and correlated single excitations ($\hat{n}_j\hat{a}^\dagger_a \hat{a}_i$).
The indices denote spatial orbitals, overbar on an index denotes $\beta$ spin, and spin-complemented interactions are implied.
For example $\hat{\tau}_{01}^{06}$ is really  $\hat{\tau}_{01}^{06} + \hat{\tau}_{\bar{0}\bar{1}}^{\bar{0}\bar{6}}$.
An interesting feature of the ansatz returned by ADAPT-VQE, Eq.~(\ref{LiHansatz}), is that the HOMO-LUMO double excitation ($e^{\hat{\tau}_{2\bar{2}}^{3\bar{3}}}$) 
	is not the first operator, but instead the third.
This is different from what one might expect if classical MP2 or CCSD amplitudes were used to order the ansatz. 
The reason is that in choosing the next operator no state energy information is used, for instance in the form of a denominator
penalizing high energy terms. Interestingly, at convergence it is not the HOMO-LUMO term or the first operator with the largest amplitude, but rather the second operator,  $e^{\hat{\tau}_{2\bar{2}}^{3\bar{6}}}$.

In Fig. \ref{fig:pes}(d-f), the dissociation curves for BeH$_2$ are shown. 
In the STO-3G basis, BeH$_2$ has 7 spatial orbitals, for a total Hilbert space dimension of 16384, 
	and a neutral molecule subspace of dimension ${7\choose 3}{7\choose 3} = 1225$. 
Unlike with LiH, UCCSD does not provide chemically accurate results across the full PES. UCCSD and ADAPT($\epsilon_1$) are comparable at smaller bond distances. Beyond $\sim$3 \AA{}, they both go above 1 kcal/mol in absolute error. 
However, still with a small fraction of the number of parameters in UCCSD, both ADAPT($\epsilon_2$) and  ADAPT($\epsilon_3$)
provide nearly exact results, with average deviations from FCI listed in Table~\ref{tbl:pes}.

\begin{table}
	\caption{Average errrors across the PES scan for the different methods assessed. Units in kcal/mol.}\label{tbl:pes}
	\begin{tabular}{l c c c c}
		\hline\hline
		&UCCSD	&ADAPT($\epsilon_1$)	&ADAPT($\epsilon_2$)	&ADAPT($\epsilon_3$)\\\hline
		LiH	&0.0480	&0.3000		&0.0058		&0.0002	\\
		BeH$_2$ &2.2384	&0.8023	&0.0907	&0.0041\\
		H$_6$	&3.7387		&4.5297	&0.3023		&0.0047 \\
		\hline\hline
	\end{tabular}
\end{table}

Now we move our focus to the H$_6$ data.
	At bond-breaking, the previous two molecules involved strong correlation between only two and four electrons, respectively.
	In order to evaluate the ability of ADAPT-VQE to converge to FCI in the presence of much stronger correlations,
	we have computed the simultaneous stretching of H$_6$, with the results presented in Fig. \ref{fig:pes}(g-i). 

	The complexity of this strongly correlated system is reflected in two obvious ways:
	1) the failure of UCCSD to achieve chemical accuracy across the curve in Fig. \ref{fig:pes}(h),
	and 2) the increased number of parameters selected in the ADAPT calculations in Fig. \ref{fig:pes}(i). 
	Despite being strongly correlated, such that higher excitation rank operators should be needed,
	both ADAPT($\epsilon_2$) and ADAPT($\epsilon_3$) provide accurate results with only one- and two-body operators.  Moreover, in the case of ADAPT($\epsilon_2$) this is achieved with fewer operators than UCCSD for most bond distances. ADAPT($\epsilon_3$) also uses fewer parameters than UCCSD up to the distance where UCCSD fails to reach chemical accuracy.

Because the algorithm is adaptive, during the course
of a chemical event (bond breaking, isomerization, etc)
the number of parameters can change abruptly, leading
to discontinuous potential energy curves. Two notable examples of this can be seen in Fig. 2(h), first at R(H-H) = 1.8 {\AA}  where ADAPT($\epsilon_1$) experiences a large jump in energy, and second at 2.5 {\AA} where ADAPT($\epsilon_2$) increases in energy.
Fig. 2(i) shows that these energy jumps correspond to sudden drops in parameter counts.

The cause of the discontinuities in the H6 data can be explained from the convergence data provided in the Supplement (see Supplement Fig. 1). 
For larger bond lengths, as additional operators are added to the ansatz, the energy flattens out before dropping substantially again. 
If the convergence criterion is too lenient, then the ADAPT-VQE optimization will abort at such ``false gradient troughs''. 
In the ADAPT($\epsilon_2$) data of Fig. 2(h) and 2(i), the jump in energy error and drop in parameter number, respectively, are caused by the 2.5 A optimization aborting at a false gradient trough, while the optimizations at other bond lengths do not.
Of course, if a tighter threshold is used (such as 0.001), the ADAPT-VQE algorithm does not prematurely abort, and ultimately yields high-accuracy results, even for this strongly correlated system.
More sophisticated convergence checks in step 6 might avoid these situations and will be one focus of future work.

\subsection{Dependence of convergence on operator ordering}\label{sec:convergence}

To demonstrate the importance of the gradient-based operator ordering chosen by ADAPT-VQE, 
	we compare it to a few alternate procedures for growing the ansatz:
		 a) Random (ijab): Randomly select from a pool of $\tau_{ij}^{ab}$, 
		   where the indices are restricted to those which do not annihilate the HF reference state.
		 b) Random (pqrs): Randomly select from a pool of $\tau_{pq}^{rs}$, 
		   where the indices are not restricted.
		 c) Lexical (ijab): Select from an ordered pool of $\tau_{ij}^{ab}$, 
		   where the indices are restricted to those which do not annihilate the HF reference state.
		 d) Lexical (pqrs): Select from an ordered pool of $\tau_{pq}^{rs}$, 
			where the indices are not restricted.

\begin{figure}
	\includegraphics[width=\linewidth]{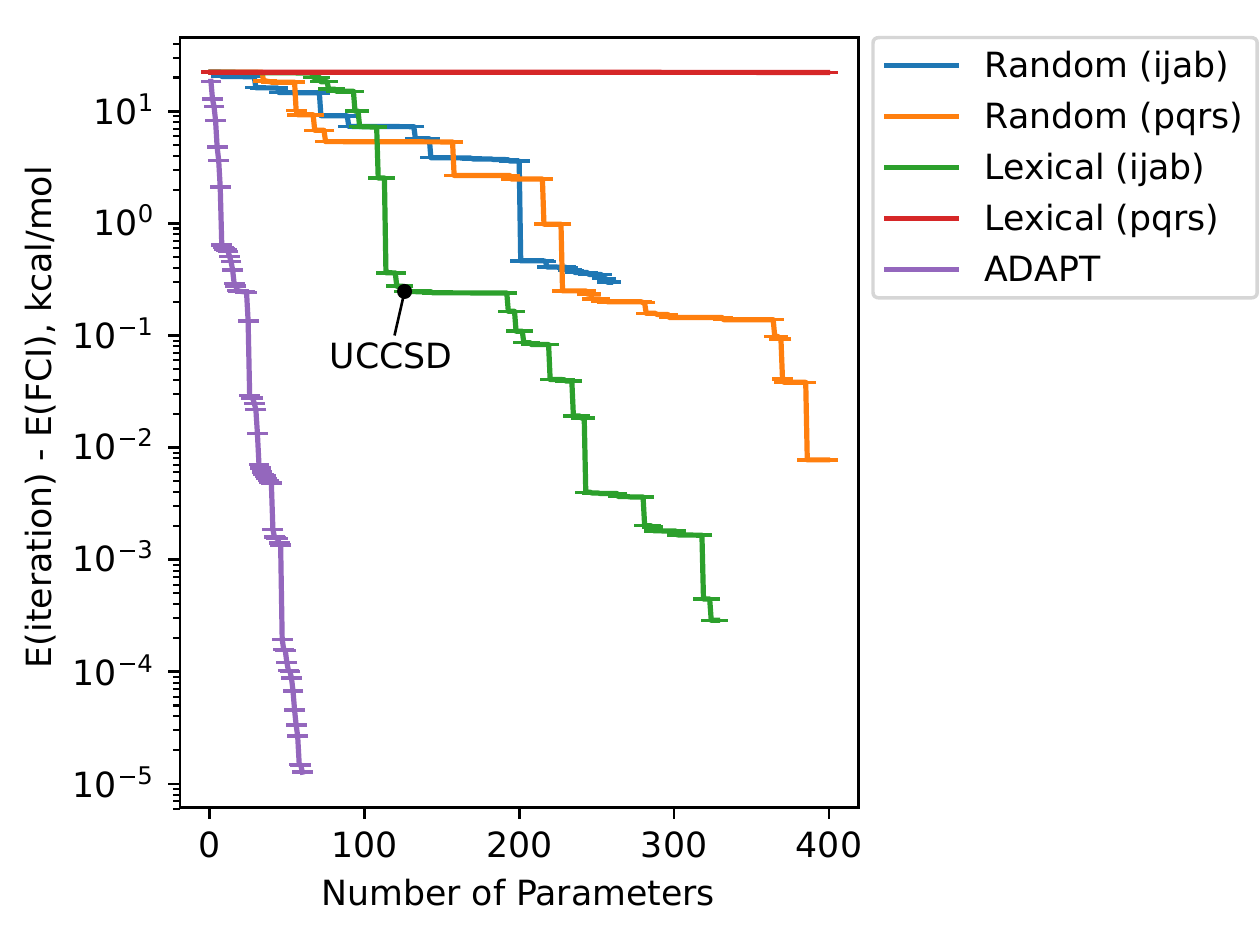}
	\caption{Convergence of the energy as a function of parameter count for BeH$_2$ with a Be-H distance of 2.39 \AA.
	The black dot indicates the UCCSD point.
	} \label{fig:convergence}
\end{figure}

In Fig.~\ref{fig:convergence}, we show the convergence of each of these orderings and compare them to ADAPT using BeH$_2$ as a typical example.
What stands out is that the ADAPT ansatz converges dramatically faster than the other four cases considered. 
While the two random-growth ans\"atze converge relatively similarly to each other regardless of whether restricted indices are used or not, 
the lexically ordered ansatz shows a clear distinction between the restricted index (singles and doubles) and un-restricted index (generalized singles and doubles)
ordering. This is due to the fact that the first operators in the ansatz involve creation operators on the occupied orbitals, 
and these do not contribute until the wavefunction has become entangled. The un-Trotterized UCCSD result is also marked for reference. 
Overall, the data in Fig. \ref{fig:convergence} demonstrate that an iterative gradient minimization algorithm yields a highly compact ansatz for a given state.

\section{Discussion}\label{sec:discussion}

An obvious metric for evaluating the performance of any simulation algorithm can be simply described as some accuracy measure vs. some cost measure.
While the accuracy measure in a simulation is often easy to define, the cost measure is more nuanced. 
For variational quantum simulations, there are two factors which largely determine the overall cost: circuit depth and number of measurements (or shot count).
Shot count is important as it determines the time to solution. 
It is possible that due to the sheer number of measurements, a particular quantum simulation becomes intractable.
However, for NISQ devices in which coherence times (and thus number of gates) are limited, circuit depth is usually the most critical cost metric, as it determines whether or not a simulation can occur at all.
By taking circuit depth as the most important cost metric to address, the original VQE has been successful by minimizing circuit depth at the cost of increased number of measurements. 
Similar to the orginal VQE, our new ADAPT-VQE algorithm seeks to further minimize the circuit depth with an increased number of measurements.

In this direction, the data clearly demonstrates that ADAPT-VQE succeeds in creating a more compact and accurate wavefunction ansatz than UCCSD.
The algorithm achieves this by systematically identifying the optimal set and ordering of operators to use in the wavefunction ansatz for a given problem. The efficiency of ADAPT-VQE makes it very promising for quantum chemistry simulations on NISQ devices, where circuit depth limitations remain a significant challenge. 

In terms of shot-count, ADAPT-VQE will likely have an increased number of measurements compared to UCCSD-based VQE due to the necessary gradient measurements. 
However, this is perhaps an easier problem to address (compared to circuit depth) as the individual runs can in principle occur simultaneously if several devices exist. 
Further, the shot count also depends on the number of iterations required for the classical optimization of the ansatz parameters. 
For strongly correlated systems where perturbation theory fails, the existing approach of using classical MP2 amplitudes to initialize the UCCSD parameters (Ref. \cite{McClean2016}) is not likely to provide much improvement in the UCCSD-based VQE.
Alternatively, each iteration of ADAPT-VQE only adds a single new parameter, with the previously optimized parameters already being initialized to rather sensible values. 
This might ultimately decrease the number of iterations needed for the VQE subroutine in ADAPT-VQE, thus decreasing the shot count (although this is not likely to fully compensate for the large number of measurements for the gradient). As hardware capabilities continue to increase, in terms of both the size and number of quantum processors available, ADAPT-VQE will offer an ideal quantum-parallel approach to performing nontrivial quantum chemistry simulations. We therefore expect this algorithm to have a strong impact on these efforts in the near term.

As the name suggests, ADAPT-VQE could be classified as one member of a family of adaptive-basis strategies that has seen success in constructing compact many-electron wavefunctions \cite{Harrison1991,Povill1992,Peris1999,Evangelista2014b,Schriber2016,Holmes2016,Xu2018} 
and single-electron wavefunctions \cite{Lyakh2010,Bischoff2013a,Laikov2011,Lu2004,Berghold2002,Lee1998,Schutt2018},
	 or as a relative of methods using sequential transformations which have been explored in the context of multireference coupled-cluster theory \cite{Evangelista2012,Evangelista2011}. Of these, the ADAPT ansatz is perhaps most closely related to the @-CC method of Lyakh and Bartlett \cite{Lyakh2010}, in which a compact set of cluster operators is iteratively determined to describe the state of interest on a classical computer. 
Our approach is distinct in that it is not only designed for a quantum computer implementation, 
	but also defined for a different wavefunction form (product of unitary operators vs. coupled cluster) and a different importance metric 
	(operator gradient of the many-electron state vs. a single electron-defined importance function, see Ref. \cite{Lyakh2010}) for determining new parameters.

An important aspect of ADAPT-VQE is that several steps of the algorithm can be implemented in multiple ways, lending it still greater versatility across a wide landscape of problems and suggesting that it should perhaps be thought of as a class of algorithms rather than a specific one. In the Results section, we already discussed a few algorithmic options, including different ways to perform the gradient-based parameter update and to determine convergence. We also mentioned the possibility of freezing early parameters at later stages of the algorithm in order to speed up the re-optimization steps. Below, we discuss a few more modifications to explore.

Although the ADAPT-VQE algorithm is notably not a perturbative approach, 
	it still has a perturbative flavor in that the suitability of the next 
	iteration's best operator only involves the interaction of that operator with the Hamiltonian.
As such, the algorithm may not be able to recognize the best quadruple excitation (for example) during one update. 
That being said, the physics described by quadruple excitations is ultimately captured after multiple iterations through the product of at least two two-body interactions. 
The consequence of this is that convergence will likely not be as fast for strongly correlated systems because the algorithm can only ``see'' two body operators at a time. 
Because only local knowledge of the FCI energy landscape is used to update the ADAPT-VQE ansatz construction, the ``true optimally compact ansatz'' is not guaranteed. 
As a result, flat energy landscapes (associated with ``false gradient troughs'') are possible. 
Further classical simulations and device implementations are needed to provide better insight into the numerical behavior.

Fortunately, however, multiple strategies can be pursued to address any possible slow convergence issues.
One possible approach would be to add a selection of three- or four-body interactions into the operator pool, 
	such that these could be inserted when needed.
Alternatively, one might imagine trying to update the ansatz with two (or more) operators in each iteration, such that the best set of operators is added.
The operator pool would still consist of only one- and two-body interactions, but higher-body interactions could be incorporated through products of operators. 
Even further, one might imagine computing the second derivative and using Hessian matrix elements to identify cooperative effects between operators in the pool.
We will explore each of these approaches in future work, with the aim of determining the fastest converging algorithm in different chemical scenarios.

In this paper, we presented ADAPT-VQE, a novel variational hybrid quantum-classical algorithm designed to achieve exact results at convergence. Unlike typical ans\"atze, which tend to be ad hoc, our approach is based on an ansatz that is determined by the system being simulated, and it features a well-defined, built-in convergence criterion. Moreover, the parameter count, and thus the gate depth, is kept to a minimum. A detailed description of the algorithm is given, and numerical examples are provided to demonstrate the performance of the ADAPT method with both weakly and strongly correlated systems. Based on these results, we find the ADAPT-VQE algorithm to be an operator- and parameter-efficient method capable of high accuracy, with controllable errors, that routinely outperforms UCCSD. Its compatibility with classical routines for compiling state preparation circuits and quantum-parallelism should make ADAPT-VQE extremely useful for simulations of molecules on both currently available and future quantum computers.

 \section*{Acknowledgements}
 
 This research was supported by the US Department of Energy (Award No. DE-SC0019199) and the National Science Foundation (Award No. 1839136). S.E.E. also acknowledges support from Award No. DE-SC0019318 from the Department of Energy.

\end{document}